\documentclass[twocolumn,aps,pra,longbibliography,superscriptaddress]{revtex4-2}
\usepackage{amsmath,amssymb}
\usepackage{graphicx}
\usepackage{hyperref}
\usepackage[normalem]{ulem}
\hypersetup{colorlinks=true, linkcolor=blue, citecolor=blue, urlcolor=blue}
\usepackage{bbold}

\begin{document}
\author{Moritz Cygorek}
\affiliation{Condensed Matter Theory, Technical University of Dortmund, 44227 Dortmund, Germany}
\author{Erik Gauger}
\affiliation{SUPA, Institute of Photonics and Quantum Sciences, Heriot-Watt University, Edinburgh EH14 4AS, United Kingdom}
\title{Time-nonlocal versus time-local long-time extrapolation of non-Markovian quantum dynamics}

\begin{abstract}
The high numerical demands for simulating non-Markovian open quantum systems motivate a line of research where
short-time dynamical maps are extrapolated to predict long-time behavior.
The transfer tensor method (TTM) has emerged as a powerful and versatile paradigm for such scenarios. It relies on a systematic construction of a converging sequence of time-nonlocal corrections to a time-constant local dynamical map.
Here, we show that the same objective can be achieved with time-local extrapolation based on the observation that time-dependent time-local dynamical maps become stationary. Surprisingly, the maps become stationary long before the open quantum system reaches its steady state. 
Comparing both approaches numerically on examples of the canonical spin-boson model with sub-ohmic, ohmic, and super-ohmic spectral density, respectively, we find that,
while both approaches eventually converge with increasing length of short-time 
propagation, our simple time-local extrapolation invariably converges at least as fast as time-nonlocal extrapolation. These results suggest that, perhaps counter-intuitively, time-nonlocality is not in fact a prerequiste for accurate and efficient long-time extrapolation of non-Markovian quantum dynamics.
\end{abstract}

\maketitle
\section{Introduction}
Understanding and mitigating the influence of the physical environment 
on quantum systems is crucial for scalable quantum technologies, which require 
high-fidelity quantum control. . 
Conventional Lindblad master equations~\cite{Lindblad,GKS} can capture the 
essential physics of open quantum systems if they are Markovian, i.e.~if memory effects induced by the environment are negligible.
However, even in well isolated systems such as superconducting 
quantum computers non-Markovian effects have been found to be 
relevant~\cite{NatCommWhite}, 
where manipulating a quantum system triggers excitations in its environment
that act back on the system at later times~\cite{deVega}.
This situation is even more drastic in condensed matter systems 
such as in semiconductor single-photon sources for quantum 
communication~\cite{PI_singlephoton,NazirReview} 
and in (bio-)chemical charge and excitation transfer 
systems~\cite{Caruso2009, Lorenzoni2024} due to their strong coupling 
to vibrations.

Accounting for the quantum dynamics of environment excitations renders 
simulating non-Markovian quantum system challenging even if the system of
interest is small. Arguments based on energy-time uncertainty and
Lieb-Robinson bounds~\cite{Plenio_LiebRobinson} further suggest that 
such simulations generally scale at least quadratically with the total 
propagation time~\cite{DnC}, while only for special environments, 
such as the spin-boson model,
numerically exact (quasi-)linear scaling methods have been 
reported~\cite{DnC,Link,SMatPI2021}.
These numerical demands have motivated the development of schemes to
predict the long-time dynamics from information obtained over much 
shorter times~\cite{TT}, which is a plausible strategy when the total 
Hamiltonian is time-independent. 

An extrapolation scheme that has attracted particular attention 
over the last decade~\cite{TT_Kananenka2016,TT_Valkunas2017,TT_Rosenbach2016,TT_InitialCor,TT_Wu2024} 
is the transfer tensor method (TTM)~\cite{TT}. There, dynamical maps
of a non-Markovian open quantum system are calculated over several time steps 
up to some cutoff time $\tau_c$ using some accurate numerical 
method. These dynamical maps are systematically decomposed into a series of 
terms, where the first term equals the dynamical map over a single time step
while the remaining terms describe temporal correlations. 
When the memory of the environment is finite, temporal correlations decrease
as a function of time differences, and thus the coefficients in this 
expansion, which are called transfer tensors, eventually become negligible.
Then, a finite number of transfer tensors are sufficient to accurately predict 
the long-time dynamics and the stationary state.
What makes the TTM particularly appealing is that it is very general, that it 
can be combined with any arbitrary short-time propagation technique~\cite{TT_Rosenbach2016}, and that it requires minimal computational effort 
beyond the short-time propagation.
Moreover, the transfer tensor decomposition also provides a succinct 
representation of non-Markovian dynamical maps, which is useful for
applications in noise characterization and reconstruction~\cite{TT_Pollock2018,TT_Hsieh2020,TT_Hsieh2022}.

On the other hand, given dynamical maps starting from
initial time $t_0$, one can also reconstruct time-dependent time-local 
dynamical maps at later times $t>t_0$ by inverting the dynamical map 
from time $t_0$ to the previous time step $t-\Delta t$~\cite{Andersson_Kraus}.
In the continuous-time limit, such time-local dynamical maps become equivalent
to time-convolutionless master equations~\cite{CanonicalLindblad}.
If the system Hamiltonian is independent of time and the memory of the environment is finite,
the time-local dynamical map (and also the corresponding master equations)
quickly become stationary on the scale of the memory time~\cite{Clark2024}.
Then long-time dynamics can be obtained simply by repeated application
of the stationary dynamical map to the reduced system density matrix.
This constitutes a time-local alternative to the time-nonlocal TTM.

In this article, we compare the performance of the TTM with the time-local 
extrapolation scheme on the examples of the spin-boson 
model with sub-ohmic, ohmic, and super-ohmic spectral densities.
To obtain short-time dynamical maps and to benchmark the long-time behavior,
we simulation open qunatum systems dynamics using the numerically exact 
process tensor matrix operator (PT-MPO) 
framework~\cite{Pollock_PRA,JP,inner_bonds} implemented 
in the ACE computer code~\cite{ACE, ACE_code}, which facilitates accurate 
simulations over very long times~\cite{JP,DnC}. 
We find that time-local extrapolation typically converges at least as 
fast as the time-nonlocal TTM. Hence, the added complexity of 
time-nonlocality brings no obvious benefit for the purpose of long-time 
extrapolation of non-Markovian quantum dynamics.

The article is structured as follows: In Sec.~\ref{sec:theory}, we summarize
the TTM and lay out the time-local extrapolation scheme. Both schemes are
then applied in Sec.~\ref{sec:results} to several examples. The results are
summarized in the Discussion Sec.~\ref{sec:discussion}.

\section{Theory\label{sec:theory}}
\subsection{Transfer tensor method}
The starting point for long-time extrapolation are the dynamical maps
$\mathcal{E}_{t_n,t_0}$, which describe the time evolution of the reduced
system density matrix $\rho_{t_n}$ from initial time $t_0$ to time $t_n$ 
on a time grid $t_n=t_0+n \Delta t$ of width $\Delta t$ via
\begin{align}
\label{eq:propagate}
\rho_{t_n} = \mathcal{E}_{t_n,t_0} \rho_{t_0}.
\end{align}
The dynamical map $\mathcal{E}_{t_n,t_0}$ can be obtained using any available 
accurate numerical open quantum simulation method by performing simulations
for a complete basis of initial reduced system density matrices $\rho_{t_0}$.

Central for the transfer tensors method (TTM)~\cite{TT} is the decomposition
of the dynamical maps as
\begin{align}
\label{eq:TTdecomposition}
T_{t_n} =&\mathcal{E}_{t_n,t_0}-\sum_{m=1}^{n-1}T_{t_{n-m}}\mathcal{E}_{t_m,t_0},
\end{align}
where $T_{t_n}$ are called transfer tensors. 
The first transfer tensor $T_{t_1}=\mathcal{E}_{t_1,t_0}$ is just the dynamical 
map over the first time step. The remaining transfer tensors describe 
corrections to the repeated application of the first dynamical map, e.g., 
$\mathcal{E}_{t_2,t_0}=\mathcal{E}_{t_1,t_0}\mathcal{E}_{t_1,t_0}+T_{t_2}$, 
which account for temporal correlations due to the finite memory.
With the decomposition in Eq.~\eqref{eq:TTdecomposition}
the propagation in Eq.~\eqref{eq:propagate} can be cast into a time-nonlocal 
form
\begin{align}
\label{eq:propagateTT}
\rho_{t_n} =& \sum_{k=0}^{n-1} T_{t_{n-k}}\rho_{t_k}.
\end{align}
Because temporal correlations generally decay with increasing time distance
$t_{n-k}$, the longer-range transfer tensors tend to become insignificant.
Assuming $T_{t_{n-k}}\approx 0$ for $t_{n-k}\ge \tau_c$, where $\tau_c$ is
a cutoff time, the sum in Eq.~\eqref{eq:propagateTT} can be restricted to
$\tau_c/\Delta t$ nonzero terms, which only requires the knowledge of
dynamical maps $\mathcal{E}_{t_n,t_0}$ up to the cutoff time $\tau_c$.  
Nevertheless, Eq.~\eqref{eq:propagateTT} can be applied indefinitely, and so
one can extrapolate the reduced system density matrix $\rho_{t_n}$ to times
$t_n\gg\tau_c$ well beyond the cutoff time.

\subsection{Time-local extrapolation}

Alternatively, one can construct time-dependent time-local dynamical
map from time $t_n$ to $t_n+\Delta t$ by inverting the dynamical map 
from the initial time $t_0$ to $t_{n}$~\cite{Andersson_Kraus}
\begin{align}
\label{eq:timelocal_map}
\mathcal{E}_{t_{n}+\Delta t,t_n}
=\mathcal{E}_{t_{n+1},t_0}\mathcal{E}^{-1}_{t_n,t_0},
\end{align}
which is possible whenever $\mathcal{E}_{t_n,t_0}$ is non-singular.
Crucial for time-local long-time extrapolation is that the 
dynamical maps $\mathcal{E}_{t_{n}+\Delta t,t_n}$, viewed as a function
of time $t_n$ for a fixed time step $\Delta t$, become stationary 
before the system dynamics itself fully equilibrates. 

This can be justified by considering a time-independent Markovian extension of 
the system, by which we
mean that the system is augmented by additional degrees of freedom such that
the dynamics of the reduced system density matrix is accurately described while
the master equation for the extended system remains time-independent. 
Trivially, one can always extend the system to include the complete environment.
Alternatives are reaction coordinate~\cite{RC}, chain mapping~\cite{TEDOPA}, and
pseudomode~\cite{Pseudomodes} methods as well as 
extensions using copies of the system at different points in time~\cite{QUAPI1}.
Assuming that the corresponding time-independent Liouvillian of the extended system has no exceptional points, i.e.~it is diagonalizable, \footnote{
If the Liouvillian has exceptional points, one can use as basis vectors $\mathbf{v}_j$ the generalized eigenvectors that bring the Liouvillian into Jordan normal form $J$. Then, 
in Eqs.~\eqref{eq:ext} and \eqref{eq:Eproj}, $e^{(-\lambda_j+i\omega_j) t}  \mathbf{v}_j$ should be replaced by the more general expression $e^{Jt} \mathbf{v}_j$, which can give rise to additional polynomial terms in the time dependence, e.g., $t^{n_j}e^{(-\lambda_j+i\omega_j) t} \mathbf{v}_j$~\cite{Minganti2018}. While the expressions become more complicated, our central arguments remain valid in the presence of Liouvillian exceptional points.
}, one can formally expand the system initial state 
$\rho^{\textrm{ext}}(0)$ in the basis of right eigenvectors $v_j$ of the Liouvillian $\mathcal{L}^\textrm{ext}$. 
The dynamics is then given by 
\begin{align}
\rho^{\textrm{ext}}(t) =&
\sum_{j=1}^{M} 
\big(\tilde{v}_j^\dagger \rho^{\textrm{ext}}(0)\big)
e^{(-\lambda_j+i\omega_j) t}  v_j,
\label{eq:ext}
\end{align}
where $\lambda_j$ and $\omega_j$ are the negative real part and the imaginary 
part, respectively,
of the $j$-th eigenvalue of $\mathcal{L}^\textrm{ext}$, 
and $\tilde{v}_j$ are the dual vectors, which are used to project
out the component of the initial density matrix along the direction of the
right eigenvectors $v_j$. $M$ is the number of relevant degrees of freedom 
required for an accurate description. 

Now, over the course of time, most degrees of freedom become irrelevant,
either because contributions with large $\lambda_j$ 
quickly decay or contributions with different frequencies $\omega_j$ 
destructively interfere. Once all but $M\approx D^2$ degrees of freedom 
become negligible, where $D$ is the system Hilbert space dimension, 
the dynamics is described by the stationary dynamical map 
\begin{align}
\label{eq:Eproj}
\mathcal{E}_{t+\Delta t,t}= \sum_{j=1}^M e^{(-\lambda_j +i\omega_j)\Delta t} 
\mathcal{P} v_j \tilde{v}^\dagger_j \mathcal{P}
\end{align}
obtained by projecting to the space of reduced system density matrices using
projection matrix $\mathcal{P}$.
In essence, Eq.~\eqref{eq:Eproj} defines stationary dynamical map 
independent of time $t$, which can be formulated
if the $M$ projected dual vectors $\tilde{v}^\dagger_j \mathcal{P}$ 
are linearly indepedent, which necessitates $M\le D^2$.

By contrast, the open quantum system itself only becomes stationary when 
excitations of all degrees of freedom except the one with Liouvillian 
eigenvalue $-\lambda_0+i\omega_0=0$ have died down, i.e.~$M=1$.

With the assumption of stationarity of dynamical maps
$\mathcal{E}_{t_{n}+\Delta t,t_n}=\mathcal{E}_s$ for times $t_n>\tau_c$, 
time-local extrapolation is established by
propagating the reduced system density matrix using
\begin{align}
\label{eq:tlprop}
\rho_{t_{n+1}}=\begin{cases}
\mathcal{E}_{t_{n}+\Delta t,t_n} \rho_{t_n}, & t_n \le \tau_c, \\
\mathcal{E}_s \rho_{t_n}, & t_n >\tau_c,
\end{cases}
\end{align}
where $\mathcal{E}_s$ is obtained by 
$\mathcal{E}_s:=\mathcal{E}_{\tau_c,\tau_c-\Delta t}$.

\subsection{Time-local master equations}
Eq.~\eqref{eq:tlprop} is sufficient for time-local extrapolation of
reduced system density matrices. Nevertheless, it is insightful to establish
the connection between time-local dynamical maps and time-local master equations.
Given time-local dynamical maps $\mathcal{E}_{t+\Delta t,t}$ for small
enough time steps $\Delta t$, one can reconstruct an effective time-local 
Liouvillian~\cite{Andersson_Kraus}
\begin{align}
\mathcal{L}_t =& \ln\big(\mathcal{E}_{t+\Delta t,t}\big)/\Delta t + 
\mathcal{O}(\Delta t),
\end{align}
which is equivalent to a time-convolutionless master equation~\cite{TCL1976,TCL1977}.
The time-local Liouvillian can be further decomposed into the 
canonical Lindblad form~\cite{CanonicalLindblad}
\begin{align}
\label{eq:Lindblad}
\mathcal{L}_t \rho = -\frac{i}{\hbar}[\tilde{H}, \rho ]
+\sum_j \gamma_j \big( L_j \rho L_j^\dagger -\frac 12(L_j^\dagger L_j \rho +
\rho L_j^\dagger L_j)\big),
\end{align}
where the effective Hamiltonian $\tilde{H}$, the Lindblad rates $\gamma_j$, and
the Lindblad operators $L_j$ all depend on time $t$. A unique decomposition
(up to degeneracies) is obtained by solving an eigenvalue problem 
if one additionally imposes the condition that all $L_j$ and $\tilde{H}$ are
traceless and the $L_j$ orthogonal to each other with respect to the inner
product $\textrm{Tr}(L_j^\dagger L_{j'})\propto \delta_{j,j'}$. 

If all rates $\gamma_j$ are positive, the dynamics is Markovian. Negative rates
describe the backflow of information from the environment to the system.
The connection between negativity of rates in the canonical Lindblad form
and various measures of non-Markovianity is discussed in 
Ref.~\cite{CanonicalLindblad}.

For the examples given in the Results Sec.~\ref{sec:results}, we use the
time-dependence of the Lindblad rates mainly as a proxy for the time-dependence
of the full dynamical maps.
To this end, the normalization of the Lindblad operators $L_j$ has to be fixed 
as changing the norm of $L_j$ affects the value of the rates $\gamma_j$.
Here, we choose the 2-norm~\footnote{By contrast, the orthogonality condition
$\textrm{Tr}(L_j^\dagger L_{j'})=\delta_{j,j'}$ would correspond to fixing
the Frobenius norm to 1.}  
of $L_j$ to be 1, 
as then there exists a state in the system Hilbert space
(the right eigenvector belonging to eigenvalue 1), which decays with
rate $\gamma_j$ by the anticommutator part of Eq.~\eqref{eq:Lindblad}.
\section{Results~\label{sec:results}}
The main goal of this article is to explore and compare numerically 
the performance of time-nonlocal and time-local 
extrapolation. The examples probe different regimes of the 
spin-boson model~\cite{Leggett}, which is defined by the Hamiltonian
\begin{align}
H=H_S+ \sum_k \bigg(\hbar\omega_k b^\dagger_k b_k
+ \hbar g_k(b^\dagger_k+b_k) \hat{O}\bigg),
\end{align}
where $H_S$ is the system Hamiltonian, $\hat{O}$ is a system operator 
describing the system-environment coupling, $\omega_k$ is the frequency 
of the $k$-th bosonic environment mode and 
$g_k$ is the corresponding coupling constant. The environment parameters
are fully defined by the spectral density 
$J(\omega)=\sum_k|g_k|^2\delta(\omega-\omega_k)$.
The spin-boson model has qualitatively different properties depending on the
behavior of $J(\omega)$ close to $\omega\approx 0$~\cite{Leggett}, where
$J(\omega)\sim \omega^s$ are classified as sub-ohmic, ohmic, or super-ohmic 
if $s<1$, $s=1$, or $s>1$, respectively.

Throughout this article, we use the ACE code~\cite{ACE,ACE_code} as 
a black-box numerically exact solver for non-Markovian open quantum systems.
The inputs are the spectral density, the initial bath temperature, 
the system Hamiltonian $H_S$, as well as a set of convergence parameters
like the time steps $\Delta t$ and the PT-MPO compression threshold $\epsilon$.
The algorithm for spin-boson models by J{\o}rgensen and Pollock~\cite{JP} is
selected. The time-dependence of reduced system density matrices are 
calculated for a full basis of system initial states to yield 
the dynamical maps $\mathcal{E}_{t_n,t_0}$, which are the starting point
of both time-local and time-nonlocal extrapolation.

\subsection{Sub-ohmic spin-boson model}
\begin{figure*}
\includegraphics[width=0.99\linewidth]{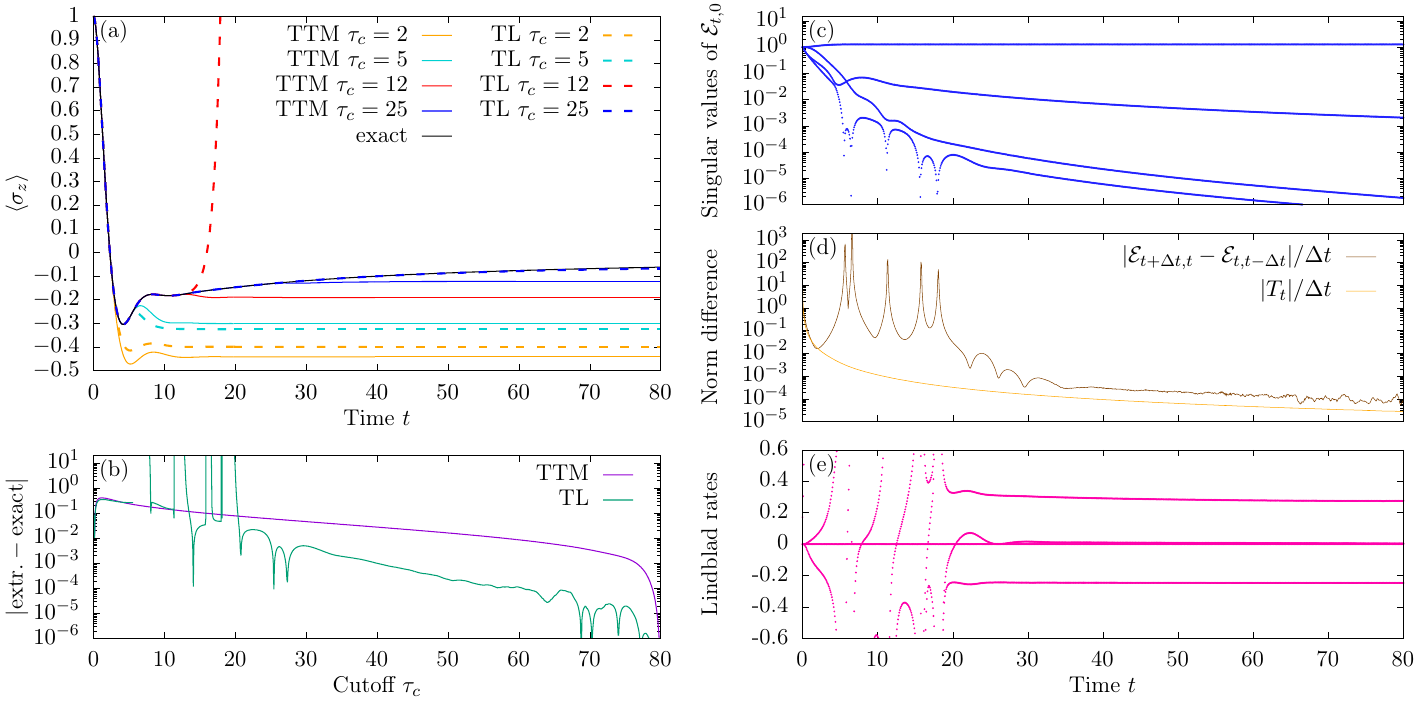}
\caption{\label{fig:sohmic}(a) Dynamics of a driven sub-ohmic spin-boson model 
including extrapolated dynamics starting from various cutoff 
times $\tau_c$ using the TTM and time-local extrapolation (TL), respectively.
(b) Difference between extrapolated (from cutoff time $\tau_c$) and 
exact value of $\langle\sigma_z\rangle$ at time $t=80$.
(c) Singular values of the dynamical map $\mathcal{E}_{t,0}$
as a function of time $t$ for $\Delta t=0.08$. 
(d) Frobenius norm of the differences between local dynamical maps over 
subsequent time steps and of the transfer tensors.
(e) Canonical Lindblad rates obtained from time-local dynamical map $\mathcal{E}_{t+\Delta t,t}$.
}
\end{figure*}
We begin our numerical analysis with a driven two-level system (TLS) 
coupled to a sub-ohmic spectral density 
since the long memory time (bath correlation functions decay algebraically) 
lead to particularly strong non-Markovian effects~\cite{Strunz_subohmic}.

We choose a driving of the form $H_S=\frac{\hbar}2 \Omega \sigma_x$, 
where the driving strength $\Omega=1$ sets the time and frequency scale.
The system is coupled to the environment via the operator
$\hat{O}=\frac{1}{2}\sigma_z$ and the sub-ohmic spectral density is taken as
$J(\omega)=2\alpha \frac{\omega^{s}}{\omega_c^{s-1}}e^{-\omega/\omega_c}$ with
$s=0.7$, $\alpha=0.2$, and $\omega_c=5\Omega$. The environment
is initially at temperature $T=0$. 

Fig.~\ref{fig:sohmic}(a) shows the time evolution of $\langle\sigma_z(t)\rangle$
starting from the maximally polarized initial state with 
$\langle\sigma_z(0)\rangle=1$ calculated using the ACE code~\cite{ACE_code}
with convergence parameters $\Delta t=0.08$ and $\epsilon=10^{-11}$.
Also shown are the results obtained when the dynamics is extrapolated 
using the TTM and the time-local (TL) extrapolation, respectively, 
starting from various cutoff times $\tau_c$.
The exact dynamics decays fast and oscillates on a short time scale 
followed by a slow monotonic decay at long time scales. 
The TTM extrapolation of $\langle \sigma_z\rangle$ 
roughly follows the tangent of the exact dynamics at
the cutoff time $\tau_c$ and then levels off to a nonzero stationary value.
As shown in Fig.~\ref{fig:sohmic}(b), the TTM-extrapolated value for 
$\langle \sigma_z(t=80)\rangle$ deviates noticeably from the exact result
unless the cutoff time $\tau_c$ approaches the time $t=80$, where the
extrapolation is evaluated.

Time-local extrapolation behaves similar to TTM extrapolation for some 
cut-off times $\tau_c$ (here at $\tau_c=2$ and $\tau_c=5$)
but may also show qualitatively different behavior for other values of $\tau_c$,
such as a divergence for some values of $\tau_c$ ($\tau_c=12$), but also
almost perfect matching to the exact results for larger cutoffs.
A clearer picture of convergence is seen in the extrapolation error as
a function of the cutoff time $\tau_c$
in Fig.~\ref{fig:sohmic}(b), which shows an abrupt transition between cutoff
times $\tau_c$ ($\lesssim 20$) for which time-local extrapolation is erratic 
and unreliable and cutoff times $\tau_c$ ($\gtrsim 20$) for which the 
time-local extrapolation works very well and is one to several orders of
magnitude more accurate that the time-nonlocal TTM.

The erratic behavior of time-local extrapolation can be explained by the
fact that the dynamical map $\mathcal{E}_{t_n,t_0}$ becomes singular 
for certain times $t_n$, as can be seen by the singular values
shown in Fig.~\ref{fig:sohmic}(c), which are obtained by
a singular value decomposition (SVD) of $\mathcal{E}_{t_n,t_0}$.
Singular dynamical maps $\mathcal{E}_{t_n,t_0}$ are know to occur when
several initial density matrices evolve to the same reduced density
matrix at some intermediate time $t_n$. This typically happens only at isolated
points in time~\cite{Andersson_Kraus}. In particular, we find that the
dynamical map is free of singularities after $\tau_c\approx 20$, which marks
the transition between erratic and well converged time-local extrapolation.

The TTM rests on the assumption that the magnitude of the transfer
tensors $|T_t|$ decay with increasing time, 
whereas the stationarity of time-dependent dynamical maps requires changes
of the dynamical maps $|\mathcal{E}_{t+\Delta t,t}-\mathcal{E}_{t,t-\Delta t}|$
to be small. Both are depicted in Fig.~\ref{fig:sohmic}(d) and are found
to decay with a similar long-time trend, yet the difference in dynamical maps
shows additional strong and wide peaks at singularities of 
$\mathcal{E}_{t+\Delta t,t}$. 
It should be noted that the error incurred on the density matrix by a single
propagation step is roughly $|(\mathcal{E}_{t+\Delta t,t}-\mathcal{E}_{s})|$, 
whereas in the TTM $(t_n-\tau_c)/\Delta t$ terms of order $|T_t|$ are 
neglected. This may explain why the TTM can be less accurate than
time-local extrapolation despite the similar long-time 
decay trend of the respective quantities.

Finally, time-dependent Lindblad rates $\gamma_j(t)$ 
extracted from the time-local dynamical maps are shown 
in Fig.~\ref{fig:sohmic}(e).
Again, singularities are found, which originate from the singularities
of the dynamical maps. After the last singularity, the rates quickly 
equilibrate with a slightly oscillatory dynamics over about $\sim 10$ 
time units. 
Notably, one of the stationary rates is negative and sizable,
indicating everlasting backflow of information from the environment to 
the system via one channel. This is physical as long as there is 
an equal or larger positive rate describing the outflow of information from
the system via another channel. The negative stationary rate indicates that
the open quantum system remains non-Markovian at all 
times~\cite{CanonicalLindblad}. 

To summarize, for the strongly non-Markovian example of a driven sub-ohmic 
spin-boson model, the time-local extrapolation scheme outperforms the 
time-nonlocal TTM as the former converges much faster 
at large cutoff times $\tau_c$. However, one should carefully check the
dynamical map obtained from short-time propagation for potential 
singularities and only apply time-local extrapolation when the dynamical 
map becomes nearly stationary.

\subsection{Ohmic and super-ohmic spin-boson model}
\begin{figure}
\includegraphics[width=0.99\linewidth]{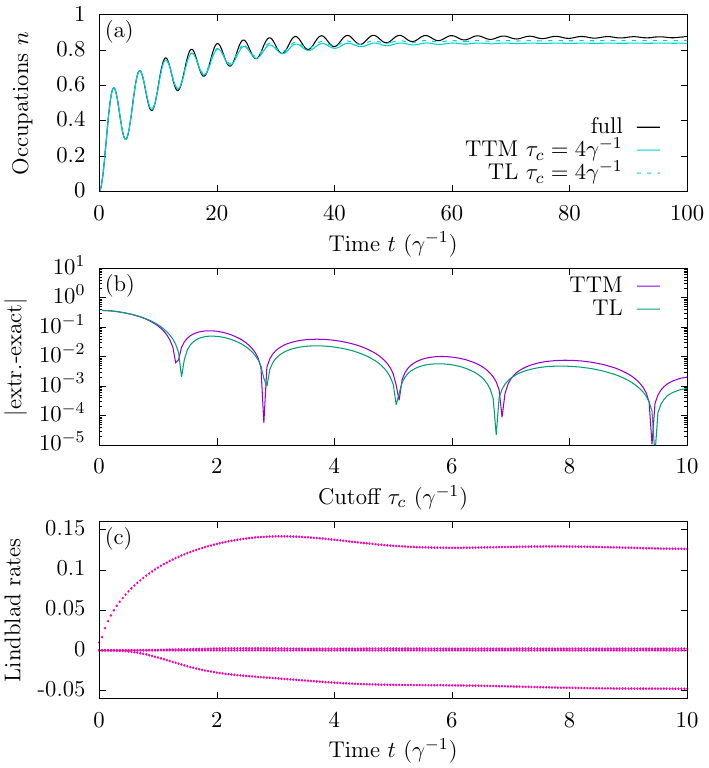}
\caption{\label{fig:DrudeLorentz}Ohmic spin-boson model with Drude-Lorentz
spectral density. (a) Dynamics with exemplary time-nonlocal and time-local
extrapolation. (b) Extrapolation error with respect to 
$\langle\sigma_z\rangle$ at time $t=100\gamma^{-1}$.
(c) Extracted time-dependent canonical Lindblad rates.
}
\end{figure}

\begin{figure}
\includegraphics[width=0.99\linewidth]{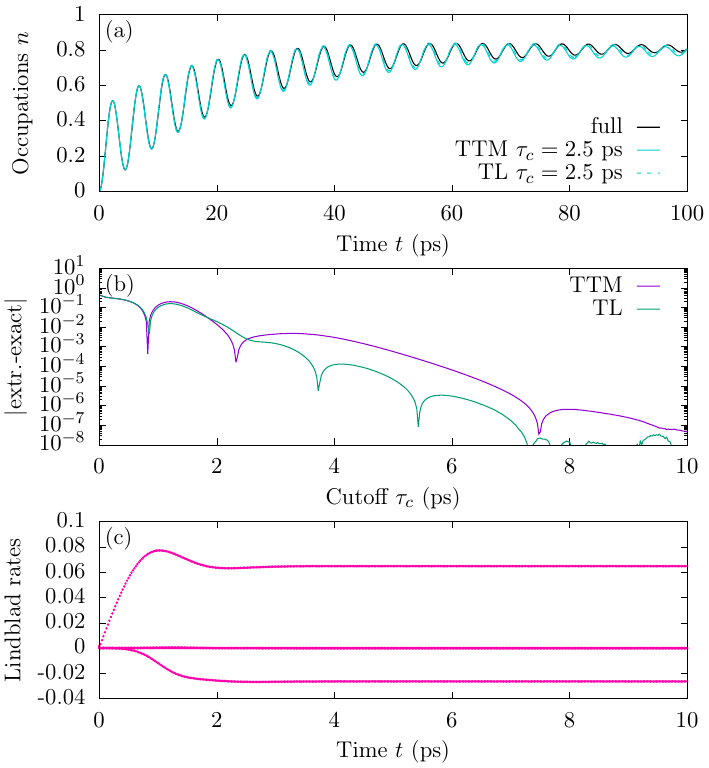}
\caption{\label{fig:superohmic}
Super-ohmic spin-boson model describing dynamics in a semiconductor quantum dot.
Panels as in Fig.~\ref{fig:DrudeLorentz}. The extrapolation error in (b) is 
evaluated with respect to reference time $t=100$ ps$^{-1}$.
}
\end{figure}

We now consider the cases of ohmic and super-ohmic spectral densities, respectively.
For the ohmic spectral density, we choose a Drude-Lorentz form
\begin{align}
J^{DL}(\omega)=\frac{2\lambda\gamma\omega}{\omega^2+\gamma^2}
\end{align}
with $\gamma=1$ sets the frequency scale and $\lambda=0.1 \gamma$ determines
the coupling strength. The system Hamiltonian is taken to be driven and biased
\begin{align}
H_S=\frac{\hbar}2 (\omega_0 \sigma_z + \Omega \sigma_x)
\end{align}
with bias $\omega_0=-\gamma$ and driving $\Omega=\gamma$. 
Convergence parameters are taken as $\Delta t=0.05/\gamma$ and $\epsilon=10^{-9}$.

For the super-ohmic example, we pick the spectral density describing the 
coupling of a semiconductor quantum dot to longitudinal acoustic 
phonons~\cite{Krummheuer2005,combine_tree}
\begin{align}
J^{QD}(\omega)=\omega^3\big( c_e e^{-\omega^2/\omega_e^2} - 
c_h e^{-\omega^2/\omega_h^2}) 
\end{align}
with constants $c_e$=0.1271 ps$^{-1}$, $c_h$=-0.0635 ps$^{-1}$, 
$\omega_e$=2.555 ps$^{-1}$, and $\omega_h$=2.938 ps$^{-1}$, which corresponds
to a quantum dot with 4 nm radius in a GaAs matrix.
We use a similar system Hamiltonian as in the ohmic case, here with detuning
$\omega_0=-1$ ps$^{-1}$ and Rabi frequency $\Omega=1$ ps$^{-1}$.

The long-time extrapolation for our models with ohmic and super-ohmic spectral 
densities are depicted in Figs.~\ref{fig:DrudeLorentz} and \ref{fig:superohmic},
respectively. The results in both cases are very similar. TTM as well as 
time-local extrapolation converge and eventually describe the dynamics well.
Both examples are much less non-Markovian than the sub-ohmic example, which
is indicated in panels (c) by the fact that the stationary negative canonical 
Lindblad 
rates are much smaller in magnitude than the respective positive rates, 
the time-dependent rates are smooth and show no singularities,
and they become nearly stationary on relatively short time scales. 
The extrapolation error shown in panel (b) is found to be almost the same for 
TTM and time-local extrapolation in the ohmic example, while for the super-ohmic
example the time-local extrapolation is again typically one to two orders of magnitude 
more accurate after the memory time of about $\tau_c=3$ ps is reached.

\section{Discussion\label{sec:discussion}}
Comparing the convergence of the time-nonlocal TTM with a time-local
extrapolation scheme utilising the eventual stationarity of 
time-dependent time-local dynamical maps, we find across several
numerical examples that including time-nonlocality  does generally not lead to better predictive power. 
On the contrary, we find examples where the extrapolation error in
time-local extrapolation is one to a few orders of magnitude smaller. 
Time-local extrapolation is only found to fail for cutoffs close to
singularities of dynamical maps. However, such singularities,
which can be directly identified from the short-term propagators by 
analyzing their singular values, tend to appear at short times below the
memory time of the environment, and thus extrapolation from cutoff times
before the last singularity seems questionable irrespective of the 
extrapolation method. Time-local extrapolation can be recommended unreservedly
for cutoff times beyond which the dynamical maps are nearly stationary, which
can be monitored by the difference between maps at subsequent time steps
$|\mathcal{E}_{t+\Delta t,t}-\mathcal{E}_{t,t-\Delta t}|$.

This finding raises the question whether one could construct alternative 
time-nonlocal approaches other than TTM that may outperform time-local 
extrapolation. The fact that time-local dynamical maps equilibrate
faster than the system itself is a promising starting point for the development
of such approaches. For example, one can imagine constructing TTMs not from
dynamical maps $\mathcal{E}_{t_n,t_0}$ starting from
the initial time $t_0$ but, e.g., from dynamical maps 
$\mathcal{E}_{\tau_c, \tau_c-n\Delta t}$ 
around the cutoff time $\tau_c$. However, this cutoff time would
have to be smaller than the memory time of the environment 
$\tau_\textrm{mem}$ in order to potentially beat time-local extrapolation. 
More generally, one could build a model for the time-dependence of the
time-local dynamical map $\mathcal{E}_{t+\Delta t,t}$ and fit it to the 
available data for $\tau_c <\tau_\textrm{mem}$. From this model one could 
then extrapolate the behavior of $\mathcal{E}_{t+\Delta t,t}$ 
towards its stationary value.
However, the complex, oscillatory, and potentially singular behavior found
for the time-dependent canonical Lindblad rates in our examples, which serve
as a proxy for the behavior of the full time-local dynamical maps, make it 
difficult to construct and justify such a model. Moreover, extrapolating to
stationary values is generally a difficult task because commonly used functional
dependencies like polynomials eventually diverge at long times, and even for
approaches that enforce stationarity, e.g. fitting of decaying exponentials,
the stationary value may depend very sensitively on the model parameters.
Thus, such approaches to
productively incorporate time-nonlocality in extrapolation schemes may be
applicable to particular problems, where the long-term behavior of the dynamical
maps is controlled, but it is difficult to construct a generic scheme that 
is accurate for very general open quantum systems problems.

Finally, we emphasize that transfer tensors can be used for a number of applications beyond long-time extrapolation, especially for tomography and analysis of open quantum systems. For example, transfer tensors describe how much the dynamics at time $t_n$ is affected by system-environment correlations at time $t_m$~\cite{Gherardini2022}. Transfer tensors can also be used to classify open quantum systems, e.g., it can be shown that dynamical maps for which all but the first transfer tensor vanish are a strictly smaller class than completely positive maps~\cite{Gherardini2022}. Furthermore, transfer tensors can also be defined for time-dependent system driving and open quantum systems with initial system-bath entanglement~\cite{TT_Pollock2018}, and they can be generalized to reproduce not only density matrix dynamics but also multi-time correlation functions~\cite{Gherardini2022}. At the same time, they can serve as an operational tool~\cite{TT_Pollock2018} for the reconstruction of a Nakajima-Zwanzig-like memory kernel describing the environment influence on the system~\cite{TT}. 
Thus, despite the fact that transfer tensors and time-local dynamical maps are both representations containing the same information as the set of dynamical maps $\mathcal{E}_{t_n,t_0}$, 
they complement each other and can provide access to different aspects of the open system. 


\acknowledgments
The authors thank Brendon Lovett for stimulating discussions. M.C. acknowledges support by the Return Program of the State of North Rhine-Westphalia as well as computation time on the LiDO3 cluster, partially funded
by the Deutsche Forschungsgemeinschaft (DFG, German Research Foundation) via
project 271512359.
%

\end{document}